# Modeling and simulation of termination resistances in superconducting cables

**Victor Zermeno[1], Philipp Krüger[1], Makoto Takayasu[2] and Francesco Grilli[1]**
[1]Karlsruhe Institute of Technology, Karlsruhe, Germany
[2]Massachusetts Institute of Technology, Cambridge, MA, USA
E-mail: victor.zermeno@kit.edu
**Abstract.** We address the problem of modeling termination resistances which are largely responsible for the uneven distribution of currents in superconducting cables. For such purpose we present three DC models. In a first model a 0D circuit-like approach considering a continuous E-J relationship is presented. A second model uses the 2D H-Formulation of Maxwell's equations, with a new contribution to the electric field term that takes into account the voltage drop due to termination resistances. A third model, based on the 3D H-Formulation of Maxwell's equations, uses a novel technique to simulate both the termination resistances and the superconducting cable within a compact framework that calculates both contributions using two non-connected domains. Advantages and disadvantages of each model are discussed. Particular applications for which a given model is best fitted are also considered. The models' predictions are in good agreement with experimental results for a stacked-tape cable composed of 4 HTS tapes. Overall, this work presents a palette of three different numerical tools for calculating the current distribution in cables composed of multiples tapes, where the termination resistance is also taken into account. The choice of one model over another depends on the particular application and on the degree of precision needed.

## 1. Introduction

The internal current repartition and the overall performance of high-current superconducting devices composed of several tapes is often seriously influenced by the presence of non-uniform resistive terminations, which add to the voltage drop individual tapes experience for a wide range of current transport amplitudes. This typically happens in laboratory-scaled cable prototypes [1] and in all those current-carrying elements where the superconductor is not long enough to compensate for the differences in the voltage drops introduced by the termination resistances. It is therefore desirable to have reliable numerical tools able to simulate termination resistances and superconductors in detail, so that solutions with optimized designs can be devised.

Several numerical models for superconductors have been developed in the past. To the best of the authors' knowledge, however, the only one that can simultaneously account for the presence of termination resistances and simulate superconducting elements in detail is the finite-element circuit-coupled model implemented in the software package Flux [2]. This model relies on the possibility of coupling "massive" conductors with circuital elements [3] and has been successfully used to simulating the influence of termination resistances in a superconducting cable [1]. However, its development and use is not trivial outside the framework of the software where it is implemented.

In this paper, we present and compare three numerical models able to account for the termination resistances in simpler terms. The models differ in terms of type and complexity, but can be easily implemented in general software packages for mathematical analysis (0D model) or in finite-element programs for the H-formulation of Maxwell's equation (2D and 3D model). We use the models to calculate the current distribution between tapes in a cable made of stacked tapes. This current distribution is determined by several factors, including the properties of the composing tapes, the self-field effects, and the presence of resistive terminations where the transport current is injected. The results of the models are compared with the experimental ones for the case of a coated conductor stacked-tape cable fabricated and characterized as reported in [4]. A schematic of such cable is shown in Figure 1. While all the models are able to predict current distributions in good agreement with experiments, they offer different degrees of details; therefore, the choice of one model over another depends on the particular application and on the degree of precision needed.

The paper is organized as follows: section 2 gives a detailed description of the three models; section 3 describes the considered test case of a stacked-tape cable; in section 4, the parameter used in the simulation are listed and their choice discussed; section 5 presents the simulation results and the comparison between the models and the experiments; finally, section 6 draws the main conclusions of this work.

## 2. Description of Models

In this work, three novel models are presented as a way to study the current distribution between tapes in cables made of stacked tapes where termination resistances are to be accounted for. All models assume quasi-static conditions with currents that vary very slowly. This is very close to the DC regime, much like the experiments are carried out in the laboratory.

### *2.1. 0D Model*

In DC conditions it is possible to consider the cable as being composed of an array of linear and non-linear resistive elements. Hence, an equivalent circuit model as shown in Figure 2 is used. As in any conventional circuit, the voltage drop across the linear resistor is modeled following Ohm's law, whereas for the nonlinear components, a power law is used for model the relation between the voltage $V$ and the current $I$ such that $V/V_c = (I/I_c)^n$, here $V_c$ is the voltage drop when the critical current $I_c$ has been reached. The exponent $n$ is used to describe how abrupt the transition from superconducting to normal state is. The $i$-th strand of the cable can be considered as a serial array of termination resistance and superconducting material such that:

$$V = V_c \left(\frac{I_i}{I_{c_i}}\right)^n + R_i\, I_i \qquad \forall i \in \{1,2,\dots,m\} \tag{1}$$

Here the sub index $i$ is used to denote the respective strand in the cable. Each equation in (1) can be rewritten as a polynomial of the form $x^n + ax - b = 0$, where $a$ and $b$ are positive constants. Using Descartes' rule of signs it is seen that this polynomial has only one positive root. In this way, the current $I_i$ can be effectively calculated from the voltage drop $V$ across the cable. Finally, the total current $I_T$ in the cable is given by:

$$I_T = \sum_{i=1}^{m} I_i \tag{2}$$

One can note that (1) converges to the linear case for values of $I_i$ well below $I_{c_i}$, while for values above $I_{c_i}$, the power law term dominates. Taking advantage of this, values for $R_i$ and $I_{c_i}$ are obtained from experimental data. Finally, since all the equations in system (1) are decoupled, they can be calculated providing a model that can rapidly be solved using standard numerical tools for solution of polynomial equations in one variable.

*2.2. 2D Model*

2D models simulate the cable's cross-section only and assume that the properties of the tapes in the cable do not vary along their length. The different materials composing the HTS tape (such as substrate or metal stabilizer) can be easily incorporated in the model, but since they are not relevant for the current distribution between the tapes of a cable, they are neglected here. Therefore, each tape is simulated as a rectangle to which superconducting material properties are assigned, as detailed in section 4.

The main issue regarding a 2D model to describe current sharing among superconducting tapes in a cable configuration is to include the contribution to the voltage drop that the termination resistances convey. A model for the cross section of a superconducting cable disregards all end effects such as termination resistances since they cannot be included while assuming infinitely long conductors. Previous strategies, like the one used in [1] dealt with this problem by considering a circuit coupling approach. Said coupling is specific to the software used in the implementation (Flux-3D) and cannot be used in a general solver for partial differential equations. On the other hand, in the present work, the voltage drop due to the termination resistances is modeled as an additional contribution to the electric field in every superconducting domain. Therefore such model still assumes an infinitely long cable while all the parameters are scaled to consider a "per unit of length" approach. This modification to the governing equation allows for its implementation in any general solver for partial differential equations without the need to have a circuit coupling artificially included.

The model presented here is a modification of the H-formulation of Maxwell equations for the low frequency regime when displacement currents are neglected as proposed in [5]–[8]. The relevant equations are:

$$\boldsymbol{\nabla} \times \boldsymbol{E} = -\frac{\partial \boldsymbol{B}}{\partial t}, \tag{3}$$

$$\boldsymbol{\nabla} \times \boldsymbol{H} = \boldsymbol{J}, \tag{4}$$

$$\boldsymbol{B} = \mu \boldsymbol{H}, \tag{5}$$

$$\boldsymbol{E} = \rho \boldsymbol{J}, \tag{6}$$

$$\boldsymbol{\nabla} \cdot \boldsymbol{B} = \boldsymbol{0}. \tag{7}$$

As explained in [9], [10], in the H-formulation, the zero divergence of the magnetic flux density as expressed in (7) is fulfilled by the time derivative in (3) and the choice of appropriate initial conditions. Equation (3) guarantees that $\boldsymbol{\nabla} \cdot \boldsymbol{B}$ remains a constant of the system. Given appropriate initial value conditions, this constant is set to 0, hence fulfilling (7). If while considering the DC case, said time derivative were to be removed, then (7) would no longer be enforced. In principle, this could be solved by introducing a gauge to (3) in order to enforce (7) in the stationary case. However, the authors have found convergence problems when dealing with the DC gauged system as described. Therefore when considering the DC case, slowly varying amplitudes will be used. This approach is described in more detail in the appendix of [11].

Combining equations (3)-(6), the magnetostatic problem can be restated as:

$$\nabla \times \rho \nabla \times \boldsymbol{H} = -\mu \frac{\partial \boldsymbol{H}}{\partial t} \tag{8}$$

As already pointed out, in the 2D case, the model neglects all end effects such as the termination resistances and considers all domains to be infinitely long. However, the voltage drop due to the termination resistances can be accounted for by separating the contributions from the superconducting tapes and the ones coming from the terminations in the computation of the electric field:

$$\boldsymbol{E} = \boldsymbol{E_{sc}} + \boldsymbol{\overline{E}_{tr}} \tag{9}$$

Here, $\boldsymbol{E_{sc}}$ is the contribution to the electric field that would normally occur in the absence of termination resistances. Complementarily, in the DC regime $\boldsymbol{\overline{E}_{tr}}$ is the averaged contribution solely due to the termination resistances. Hence, $\boldsymbol{E}$ is the electric field in the equations (3) and (6). For the DC case:

$$\boldsymbol{\overline{E}_{tr}} = \overline{R}_i \, I_i \tag{10}$$

Here, $I_i$ the current in the $i$-$th$ superconducting tape and $\overline{R}_i$ is the scaled resistance (in $\Omega/\mathrm{m}$) of the termination at its ends. The scaling in the values for $\overline{R}_i$ allows considering a “per unit of length” model. Combining equations (3)-(6) again but considering (9) and (10), and after some simple substitutions the governing equation becomes:

$$\nabla \times (\rho \nabla \times \boldsymbol{H} + \overline{R}_i \, I_i) = -\mu \frac{\partial \boldsymbol{H}}{\partial t} \tag{11}$$

For the superconducting domains, a power law $\boldsymbol{E} - \boldsymbol{J}$ characteristic is used such that:

$$\boldsymbol{E} = \rho \, \boldsymbol{J} \text{ where } \rho = \frac{E_c}{J_c} \left| \frac{J}{J_c} \right|^{n-1} \tag{12}$$

Here, $E_c$ is the critical electric field defined as $1\mu V/cm$, $J_c$ is the critical current density whose value can depend on the magnitude and direction of the local magnetic flux density. The exponent $n$, just like in the previous sections is used to describe how abrupt the transition from superconducting to normal state is. Use of (11) allows using a 2D model that geometrically depicts only the superconducting and insulating regions. In the insulating domains the term $I_i$ is set to be equal to zero. For a cable with $m$ tapes, whose *i-th* superconducting has with cross-section $C_i$, $I_i$ is given by:

$$I_i = \int_{C_i} \boldsymbol{\nabla} \times \boldsymbol{H}\, dA \quad \forall i \in \{1,2,\dots,m\} \tag{13}$$

To account for the coupling of the tapes in the cable, a non-local boundary condition in the form of an integral constraint is used to set the total current $I_T$ flowing in the cable as:

$$I_T = \sum_{i=1}^{m} I_i \tag{14}$$

Just as in [9], [10] a zero initial value for the magnetic field is used. Values for $\overline{\mathrm{R}}_\mathrm{i}$ are obtained by scaling the values of $\mathrm{R_i}$ previously calculated from experimental data as described in the previous section. Finally, equation (11) along with boundary conditions given by (13) and (14) are solved numerically using the finite element method.

*2.3. 3D Model*

A complete 3D model considering the termination resistance in superconducting cables would require simulating the actual layout of the cable at hand. This means making a very large model to consider the effect of the terminations in the full length cable. Numerical solution of such model is in principle not practical for cables with lengths beyond a few centimeters. In this work we present a strategy that allows modeling only a small section of the cable in 3D while considering the effect of the termination resistances at its ends. The main idea is to build a minimum unit cell for the superconducting region of the cable. In straight cables this accounts for an arbitrarily short section (see Fig. 5) while for twisted cables the minimum unit cell has a length equal to one twist pitch (see Fig. 7.12 in [12]). Similarly to the 2D model, only the superconducting layers and the surrounding insulating domain are simulated in this case, as detailed in section 4.

The 3D model for the H-formulation of Maxwell equations has been described in detail in [10], [13], [14] and its governing equation corresponds to (8) in the previous section. The main contribution of this work to modeling transport current in superconductors in 3D is that here, two non-connected domains are used instead of one as shown in Figure 5. This allows modeling both the termination resistances and the superconducting cable separately. This requires only a minimal geometrical layout with a small mesh in the discretization and a consequently low number of degrees of freedom. This modeling strategy is presented in detail in chapter 7 of [12].

Just like in the 2D case described above, a net current in the cable is set by means of a non-local boundary condition. Said condition is set so that the net current $I_T$ is applied to the resistive terminations in a sole integral constrain, hence replicating the electric coupling of the conductors. Figure 5 shows a straight cable parallel to the z axis. In the current transport case, the current density $\boldsymbol{J} = \boldsymbol{\nabla} \times \boldsymbol{H}$ has only one component and it is also parallel to the z axis. Considering such layout, the integral constraint can then be written as:

$$I_T = \int_{R_B} \boldsymbol{\nabla} \times \boldsymbol{H}\, dxdy \tag{15}$$

Here, the integration domain $R_B$, is the bottom surface of all the resistive terminations as shown in Figure 5. The net current in each termination is then mapped to its corresponding superconducting tapes via a set of integral constraints of the form:

$$\int_{R_{Ti}} \boldsymbol{\nabla} \times \boldsymbol{H}\, dxdy = \int_{SC_{Bi}} \boldsymbol{\nabla} \times \boldsymbol{H}\, dxdy \tag{16}$$

where the integration domains $R_{Ti}$ and $SC_{Bi}$ are respectively the $i$-th top surface of the resistive terminations and the $i$-th bottom surface of the superconducting tapes as shown in Figure 5. In this way the current is allowed to be redistributed from the terminations to the tapes.

Analogous to the 2D case discussed in the previous section, the resistivity of the terminations is obtained by scaling the value of $R_i$ previously calculated from experimental data as described before. In order to consider a compact model, the length of the geometrical model for the superconducting cable was effectively scaled down by a factor of 100. To compensate for this reduction, the resistivity of the terminations was scaled down by the same factor. This allowed simulating the superconducting cable along with its resistive terminations within a compact framework and without using simplifying assumptions.

### *2.4. Comparison of models*

The main purpose of this work is to present three different models for simulating the effect of the termination resistance in superconducting cables. Their applicability will depend on the specific scenario under consideration, the accuracy needed and the computing effort allowed for finding a solution.

The 0D model can be used when considering cables in current transport conditions. However, this model does not allow accounting for the self-field effects. It assumes a constant critical current per superconducting tape, although it is possible to assign different critical currents values to different tapes. For applications in the low current regime, such analysis can provide precise information regarding the current repartition among different strands as the self-field effects are to kick in only after larger current fractions are reached. Furthermore, the critical current in each superconducting tape can be tuned so that it reflects the current degradation due to the externally applied field at a given current amplitude. As mentioned before, simulation using this model is reduced to solving a system of decoupled polynomial equations. This can be done almost instantaneously using standard numerical tools, such as Mathematica or Matlab. Therefore, this model is valuable when a fast study of the impact the termination resistances have in current sharing is needed.

The 2D model can account for self-field effects and therefore it can be used when it is necessary to consider the critical current degradation due to either self or externally applied magnetic field. However, the case of twisted conductors with both transport current and externally applied field cannot be considered with this model as such configuration provides a true 3D electromagnetic problem. The 2D model is solved using the finite element method in 2D and therefore it requires a larger time both for implementation and for computing of simulations when compared to the 0D model.

Like the 2D model, the 3D model can also account for the self-field effects. Unlike the 0D and 2D models it can also consider twisted conductors with both transport current and externally applied magnetic field. So, the 3D model provides the most complete approach. This model is also solved using the finite element method and, for the case of non-twisted conductors treated here, requires a similar time for implementation and computing of simulations than the 2D model. In this work, for consistency with the other models presented, the 3D model will only be used for untwisted cables carrying a transport current. Study of twisted conductors with both transport current and externally applied magnetic field requires a more dedicated analysis that is out of the scope of this work and will be considered for future investigations.

## 3. Test case

In order to test the models presented in the previous section, a 60 cm long stacked-tape cable composed of 4 stacked superconducting tapes as reported in [4] was considered, each tape being 4 mm wide. The cable has no separation gap between the tapes, hence a distance of 100 μm between the superconducting layers is assumed to account for the non-superconducting materials in the tapes (substrate, stabilization, etc.). In the present work, the case of a non-twisted cable was considered. The stacked-tape cable has resistive terminations at both ends. The test case consisted of applying a slow current ramp to the cable without background field until a net transport current of 275 A is reached. The ramp was calculated so that a current rate of 0.68 A/s was applied to the cable. Therefore when considering the DC case, slowly varying amplitudes will be used. This approach is described in more detail in the appendix of [11].

For reference, the individual tapes in the cable are tagged with the numbers 1 through 4 from top to bottom. Figure 3 shows the experimental data for the intended test case. In Figure 3 one can note that for a cable current of 200 A the current in tape 1 decreases with increasing total current. A similar effect, but of lower magnitude is also seen in tapes 3 and 4. This effect is also noted in the I-V curve in Figure 4. Here the effect of the field dependence is clear as for a higher total current, the individual critical currents are reduced significantly. Furthermore, tapes 1 and 4 (top and bottom of cable), being the ones exposed to the highest magnetic fields show a more pronounced field dependence than tapes 2 and 3 (center of cable).

## 4. Models' parameters

For the 0D model and following the description at the end of section 2.1, values for the termination resistances and the critical currents of each tape were obtained from experimental data by considering the low current and high current regimes respectively. All parameters used in the 0D model are shown in Table **1**. As described before, the 2D model was set to be a "per unit of length" model. The values of the termination resistances in Table **1** were then scaled accordingly. In the same way, for the 3D model that considers a cable of scaled length, the values of the termination resistances were scaled too. In this way all three models can be fairly compared.

Characterization data for the dependence of the critical current with respect to the applied magnetic field for amplitudes close to the self-field of the tapes used in the experimental test was not available. Therefore, a general elliptic relationship was used, similar to the one presented in [15], where characterization of 4 mm wide tapes was made for a different cable layout. Although this relationship was obtained from characterization of other tapes, the overall idea is to account for the effect the magnetic field has on decreasing the critical current of the tape while in a cable configuration. The corresponding relationship is as follows:

$$J_c(B_{\parallel}, B_{\perp}) = \frac{s_i J_{C0}}{\left(1 + \frac{\sqrt{(k\, B_{\parallel})^2 + B_{\perp}^2}}{B_c}\right)^b} \tag{17}$$

here $J_{C0} = 21.218\, GA/m^2$, $k = 0.275$, $b = 0.6$, $B_c = 32.5\, mT$ and $B_{\parallel}$ and $B_{\perp}$ are respectively the parallel and perpendicular components of the magnetic flux density. The parameter $s_i$ is used as a tuning factor to account for the different values of critical current in each tape of the cable. The corresponding values for $s_1, s_2, s_3$ and $s_4$ are respectively 1, 0.9904, 0.9604 and 0.8643. These factors are used to provide different values to the tapes' critical currents, taking into account the values shown in Table 1 and adjustments to consider tape non-uniformities or degradation due to damage during manipulation.

For numerical stability and computational speed purposes, both the 2D and the 3D models consider the domain surrounding the superconducting tapes to be an insulating material with a resistivity value of $1\,\Omega \cdot m$.

## 5. Results and discussion

Figure 6 shows the individual current in each tape as a function of the total current in the cable as calculated by the 0D model and compared with the experimental results. As expected, good agreement is shown for the low current regime where the voltage drop across the cable is largely dominated by the linear termination resistances. Good agreement is obtained for the very large current regime, here the voltage drop across the cable is largely dominated by the power law $V - I$ characteristic of the superconductors and therefore the agreement is seen when the tapes approach their critical current values. In the medium current region, only a fair qualitative agreement was achieved. In this respect, it is important to observe the behavior of the current distributions measured experimentally as shown in Figure 3. For cable currents below 150 A, the current distribution is largely dominated by the linear resistance of the terminations. However with the increasing current, the nonlinear resistance of the superconducting tapes has a larger effect in the overall current sharing. After a cable current of 200 A, the current in tape 1 decreases with the increasing total current in the cable. A similar effect, but of lower magnitude is also seen in tapes 3 and 4 at total cable currents of about 220 A.

A better understanding of this phenomenon can be obtained from Figure 4. After the current in tape 1 passes 65 A, the voltage drop keeps on increasing even when the current decreases back to 65 A. Again, a similar but smaller effect is also seen in tapes 3 and 4. This current "return" is a clear sign of field dependence on the current capacity of the tapes. As the total cable current grows, the local magnetic field in the tapes decreases the critical current, and hence increases the voltage drop across them. This change in voltage drop forces the current to re-distribute to a tape whose voltage drop is still largely dominated by the linear resistance of its terminations (tape 2). This behavior continues until the current in all the tapes gets closer to their critical value, when the voltage drop given by the resistive terminations is much smaller than the one given by the superconducting tapes. Therefore, the qualitative agreement shown in Figure 6 is deemed to be very good considering that the 0D model does not take into account any field dependence.

Simulation results for the 2D and 3D models are shown respectively in Figure 7 and Figure 8. Both models provide a much better agreement with experimental data for the whole range of cable currents than the one obtained with the 0D model. The current reduction in tapes 1, 3 and 4 observed in the mid current range is well reproduced by both models. This was expected as these models take into account the field dependence of the critical current density, hence providing a clear advantage with respect to the 0D model. However, this was obtained at higher implementation and computational times.

Overall, the 3D model provided a better agreement with respect to experimental data than the 2D model. This difference is due the way in which the resistance of the terminations is taken into account by each model. While the 2D model adds a contribution to the electric field in the form of the $\overline{\boldsymbol{E}}_{\boldsymbol{tr}}$ variable, the 3D model uses a disconnected domain for the terminations and then links the currents by means of integral constraints. Therefore the contribution to the voltage drop across the terminations given by both methods will be subject to different numerical tolerances and errors. Further studies should be carried out regarding this issue to improve the convergence of the 2D model.

The remaining difference between the 3D model and the experimental data can be easily understood considering that a general expression for $J_c(B_{\parallel}, B_{\perp})$ was used and was not one explicitly derived for the tapes used in the experiment. Finally, other factors such as tape non-uniformity, degradation by handling and experimental error when measuring the currents in the individual tapes using Hall probes should also be considered as sources of the discrepancies.

Values for the critical current in each tape in the cable were calculated in each modeling strategy. As explained in section 4, for the 0D model, these values are obtained by fitting with experimental data in the high current regime. For the 2D and 3D models, critical currents were calculated following the method described in [11]. In brief, this method uses the averaged electric field as a measure of the critical voltage per unit length, for slow ramps like the ones used here; this is considered a good approximation as induction currents are negligible. The critical current estimates obtained by each model are shown in Figure 9. Although some dispersion is observed, the estimated values are within 3.5% of experimentally observed values (average error 1.4%). In particular, and as described above, the small difference between the estimates given by the 2D and 3D models are also due to the way in which the resistance of the terminations is taken into account. Overall, all estimated values are in very good agreement with experimental data; this considering that the error in measuring the currents in the individual tapes using Hall probes is in a similar range.

## 6. Conclusion

In this work, three different models for studying the effect of termination resistances on the performance of a superconducting cable were presented. The models were used to study a stacked-tape cable and simulate the internal current distribution among tapes in DC conditions.

A circuit-like 0D model provided good agreement with experimental data for low and very high cable currents. However, only fair qualitative agreement in the medium and high current ranges, where the current distribution is expected to be largely affected by the critical current dependence on the local magnetic field. Considering the ease in its implementation and computational speed, this model is valuable when a fast study is needed, especially in the low current regime.

The 2D and 3D models use an elliptic relationship to account for the dependence of the critical current with respect to the magnetic field. Being based on finite element analysis they require larger implementation and computing times than the 0D model. However, both 2D and 3D models provide a much better agreement in a wider range of current amplitudes. In particular they can account for the current return in tapes 1, 3 and 4 observed in the mid current range. In general the 3D model gives a slightly better agreement than the 2D with experimental results. This difference is expected to be due to the way the resistance of the terminations is implemented in both models numerically. Since the 2D and 3D models provide the best agreement, they are intended for applications where a detailed study of the terminations' resistance for a wide range of currents is necessary.

Finally, despite being inherently different, all models were able to estimate the critical current of each tape in the cable configuration within a small error with respect to the experimentally measured values.

## 7. Acknowledgements

Three authors (VZ, FG and PK) thank the Helmholtz University Young Investigator Group Grant VH-NG-617 for partially funding this work.

**Figures and Tables**

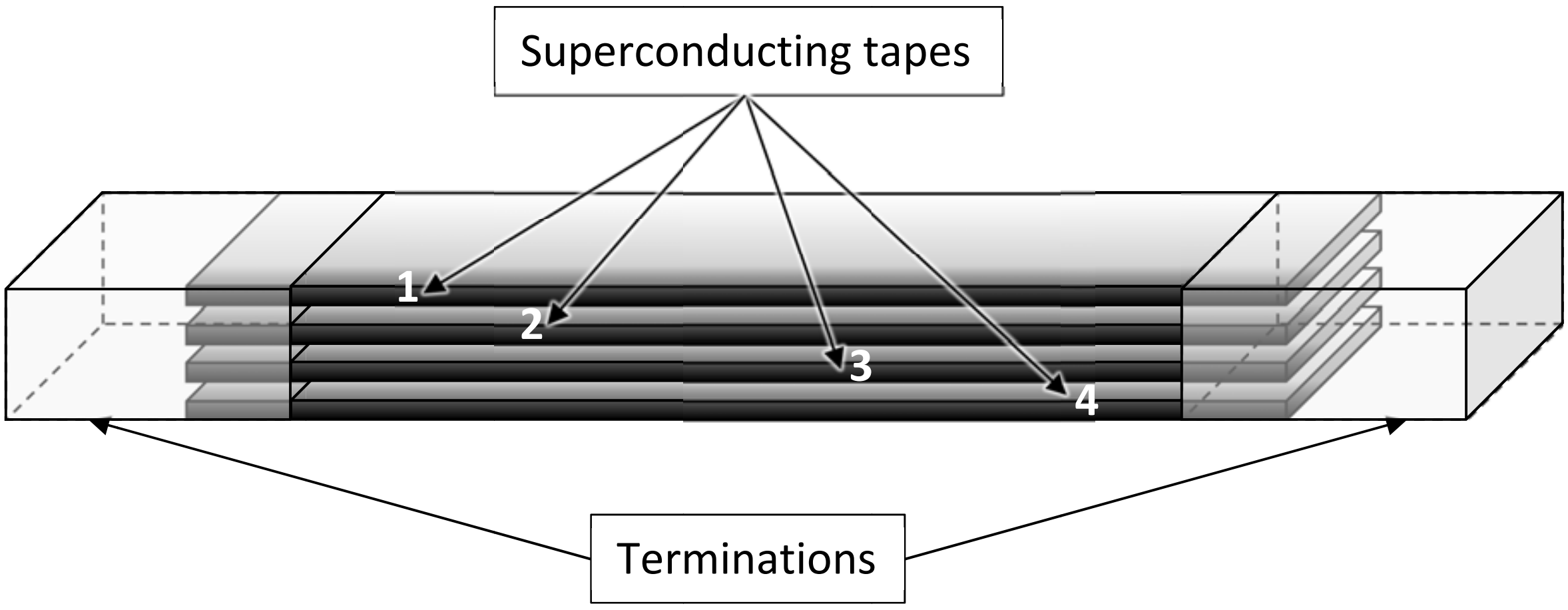

Figure 1 Structure of a stacked-tape cable (not to scale) composed of 4 HTS tapes. The tapes are connected at both ends by terminations. For reference, numbers 1 through 4 are used as tags for the individual tapes in the cable.

.

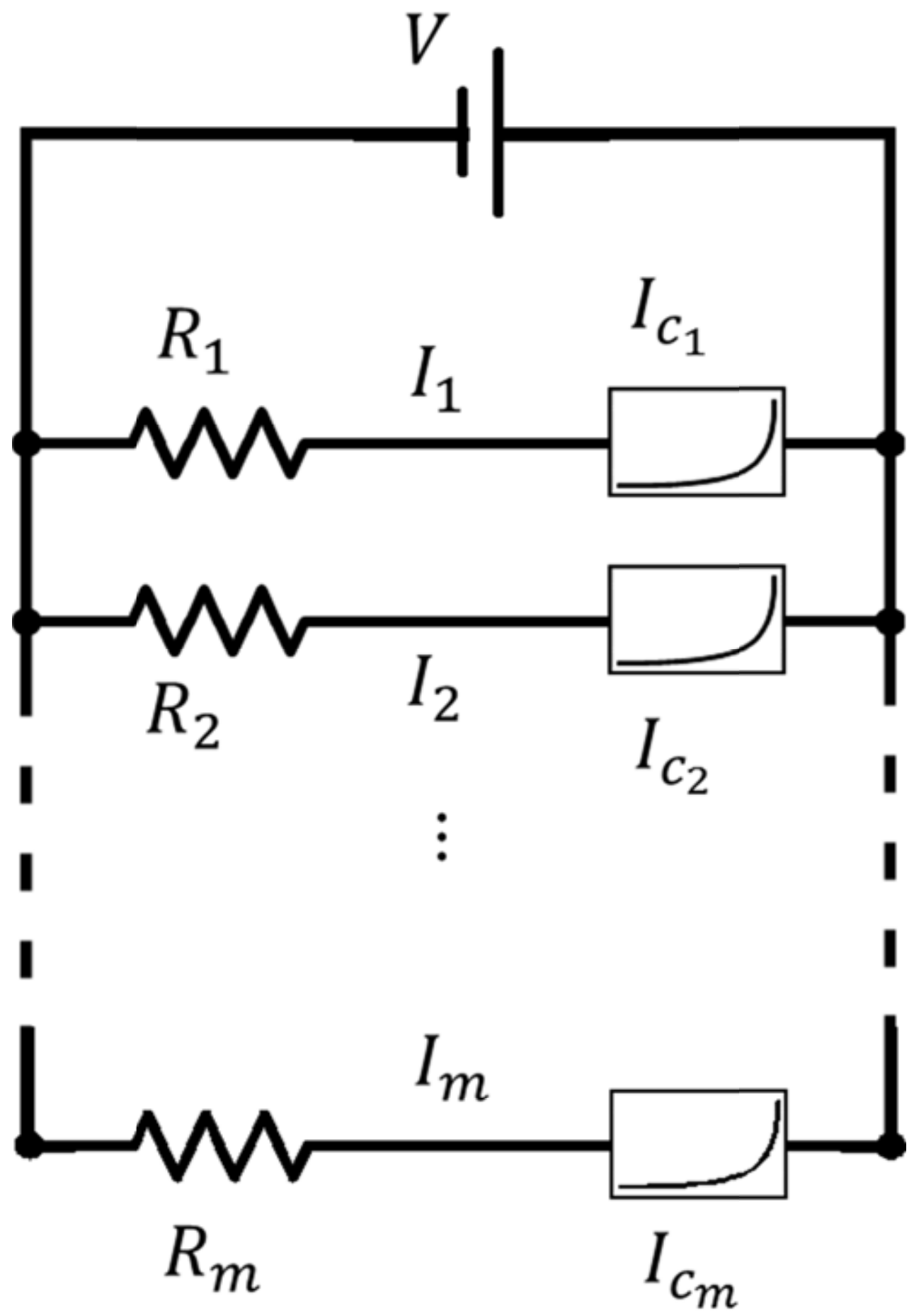

Figure 2 Diagram of the equivalent circuit model used in the 0D model.

.

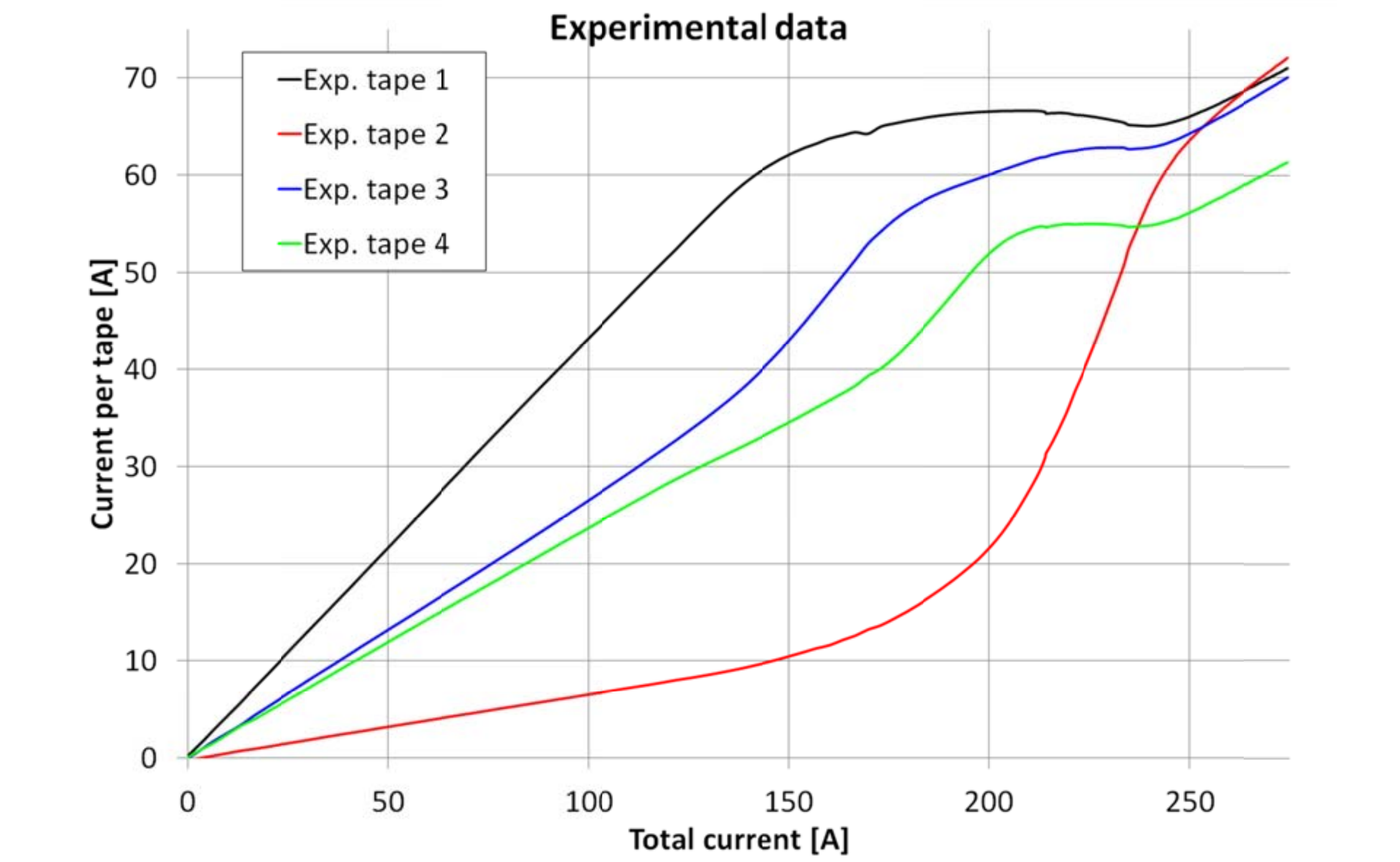

Figure 3 Experimental data: currents in individual tapes as a function of the total current in the cable.

.

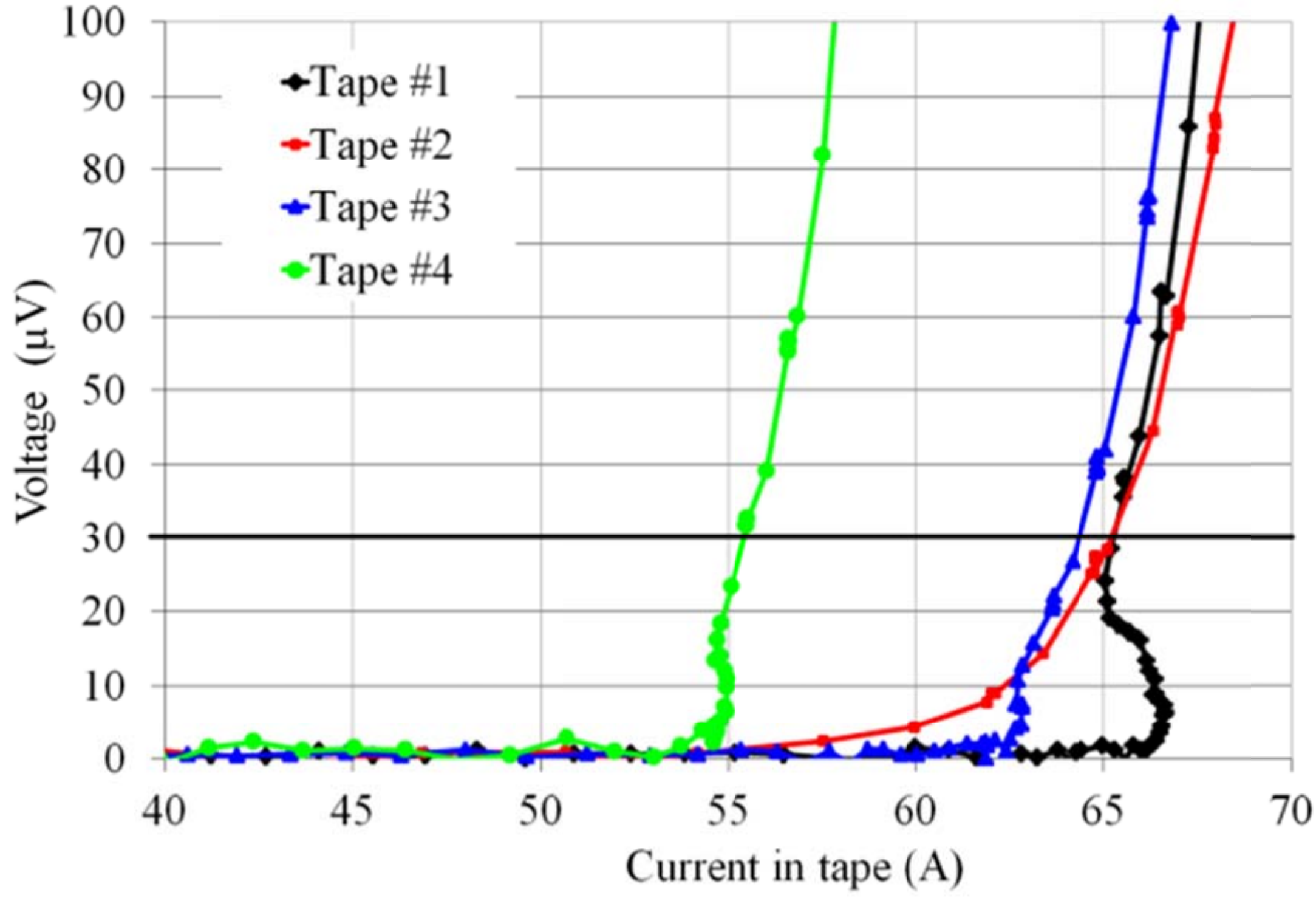

Figure 4 Experimental data: I-V curves for the individual tapes. Voltage taps were located on each tape with a separation distance of 30 cm, therefore the reference critical voltage denoted by the thick gridline is 30 μV in accordance with a 1μV/cm criteria.

.

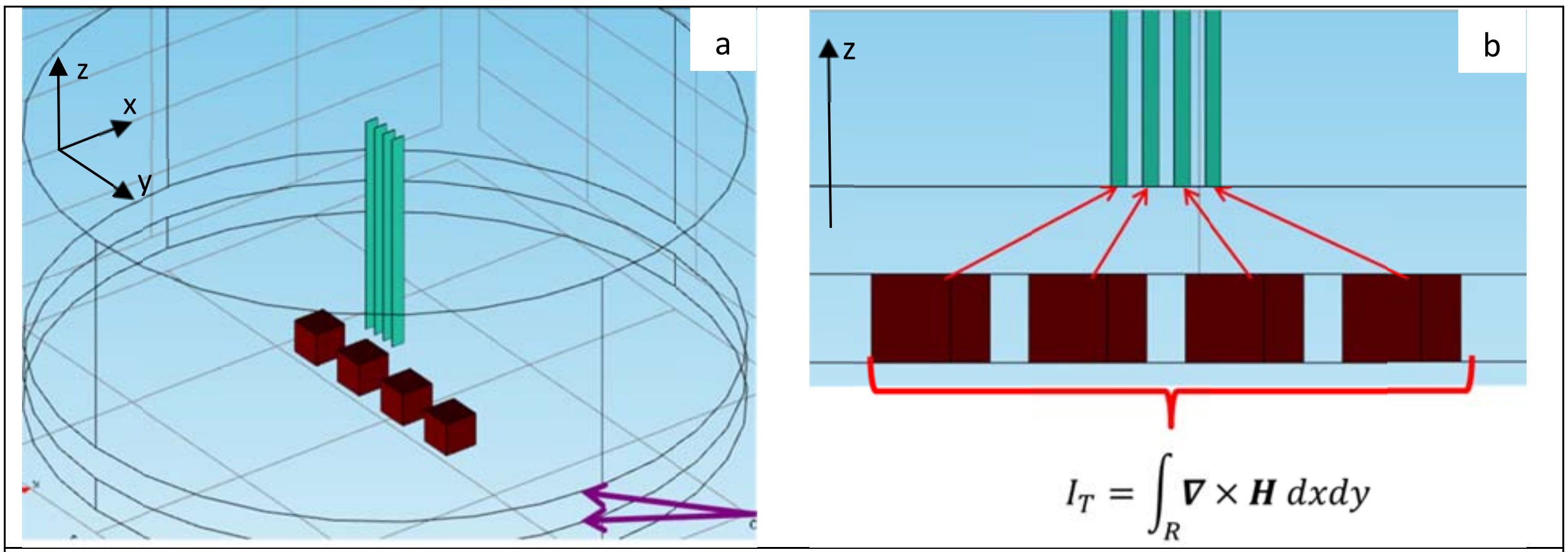

Figure 5 Computational domains used for the 3D model of the cable (not to scale). The cable is represented as the 4 superconducting tapes (green) and the terminations by the red cubes below. The cable is assumed to be parallel to the $z$ axis. The purple arrows point to the separation gap between the two non-connected domains (a). Use of integral constraints to impose a net current on the cable and the net current in each termination is mapped to its corresponding superconducting tape (b).

.

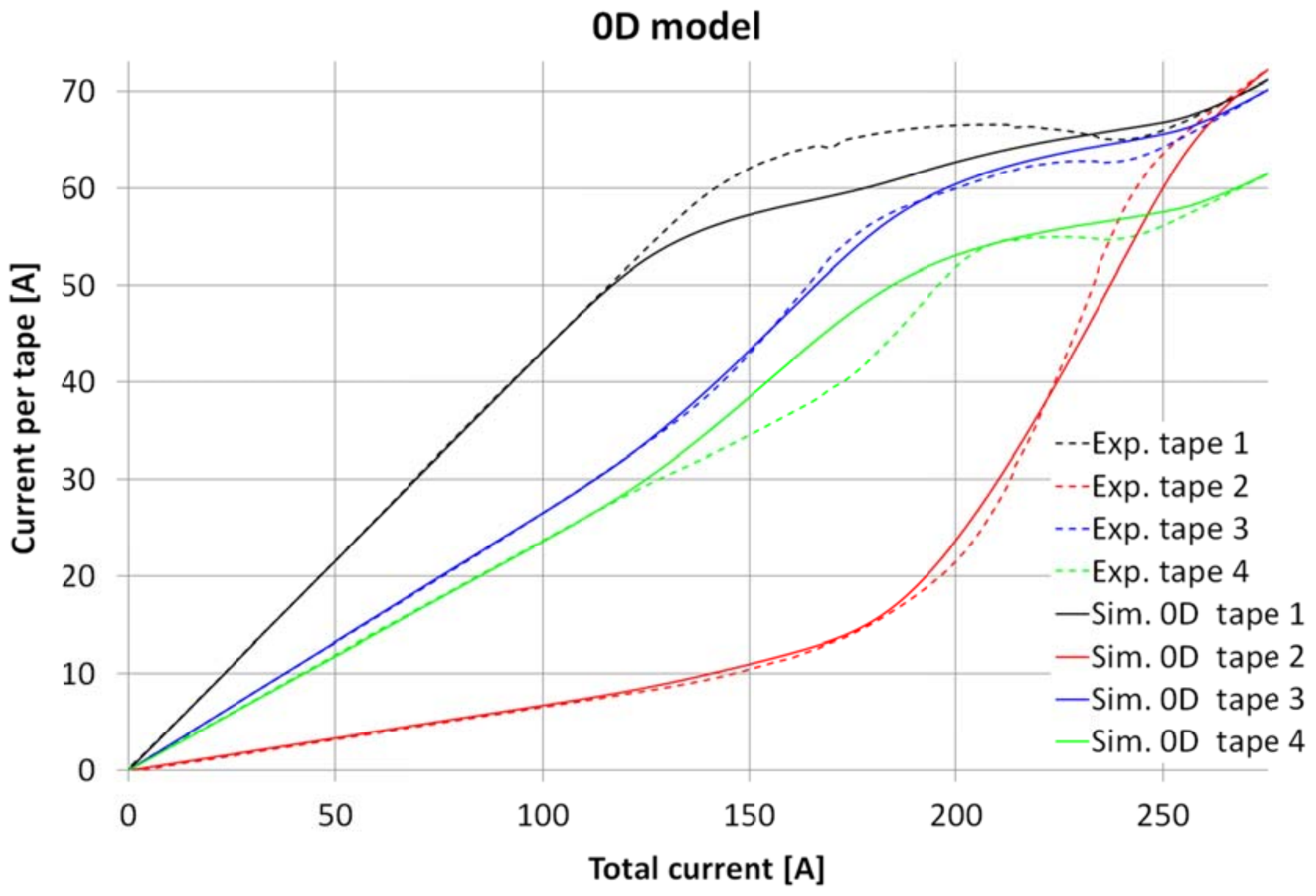

Figure 6. Current per tape vs. total current in the cable as computed by the 0D model. Dashed lines denote the experimental data and solid lines, the simulation results for the respective tapes.

.

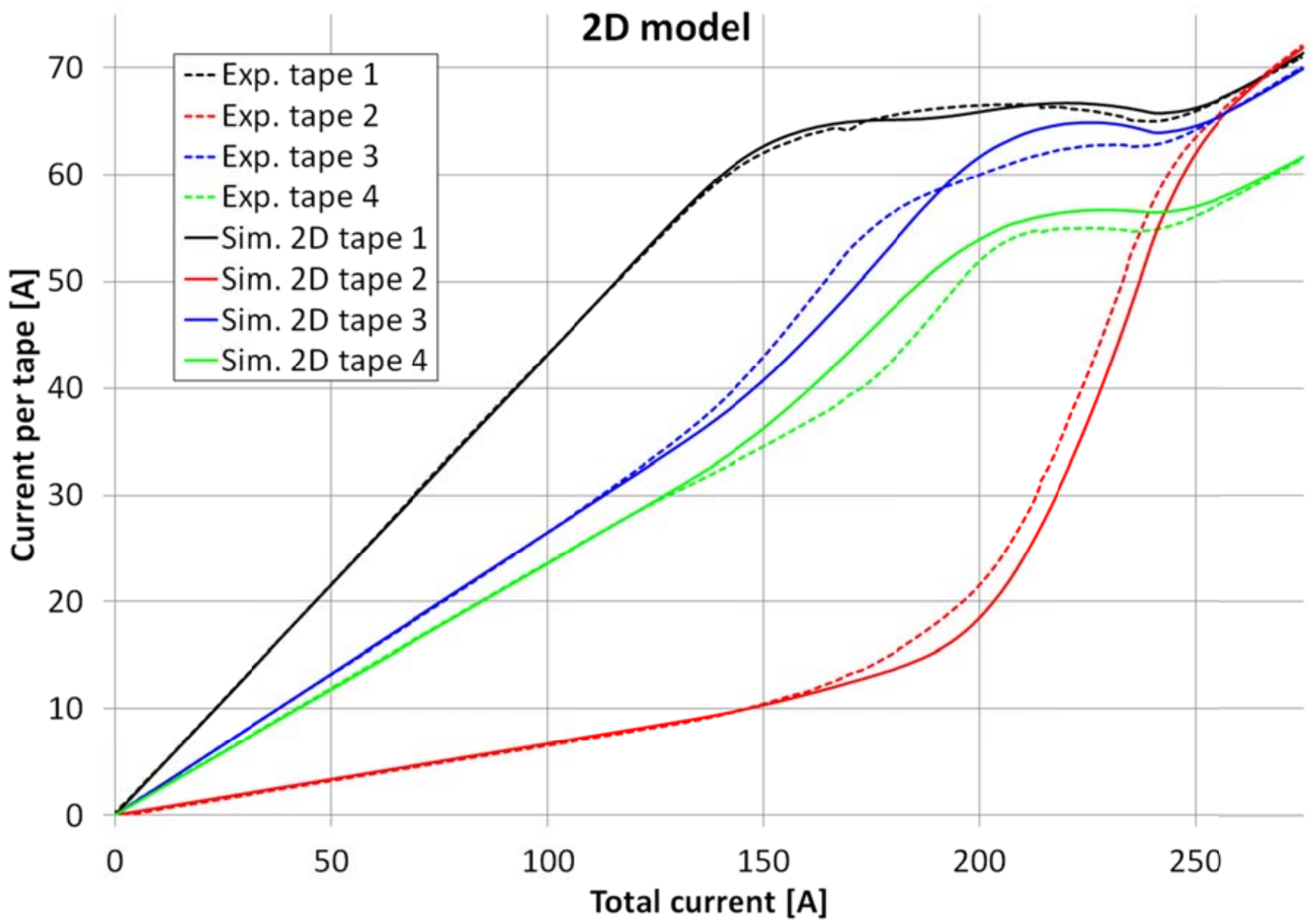

Figure 7. Current per tape vs. total current in the cable as computed by the 2D model. Dashed lines denote the experimental data and solid lines, the simulation results for the respective tapes.

.

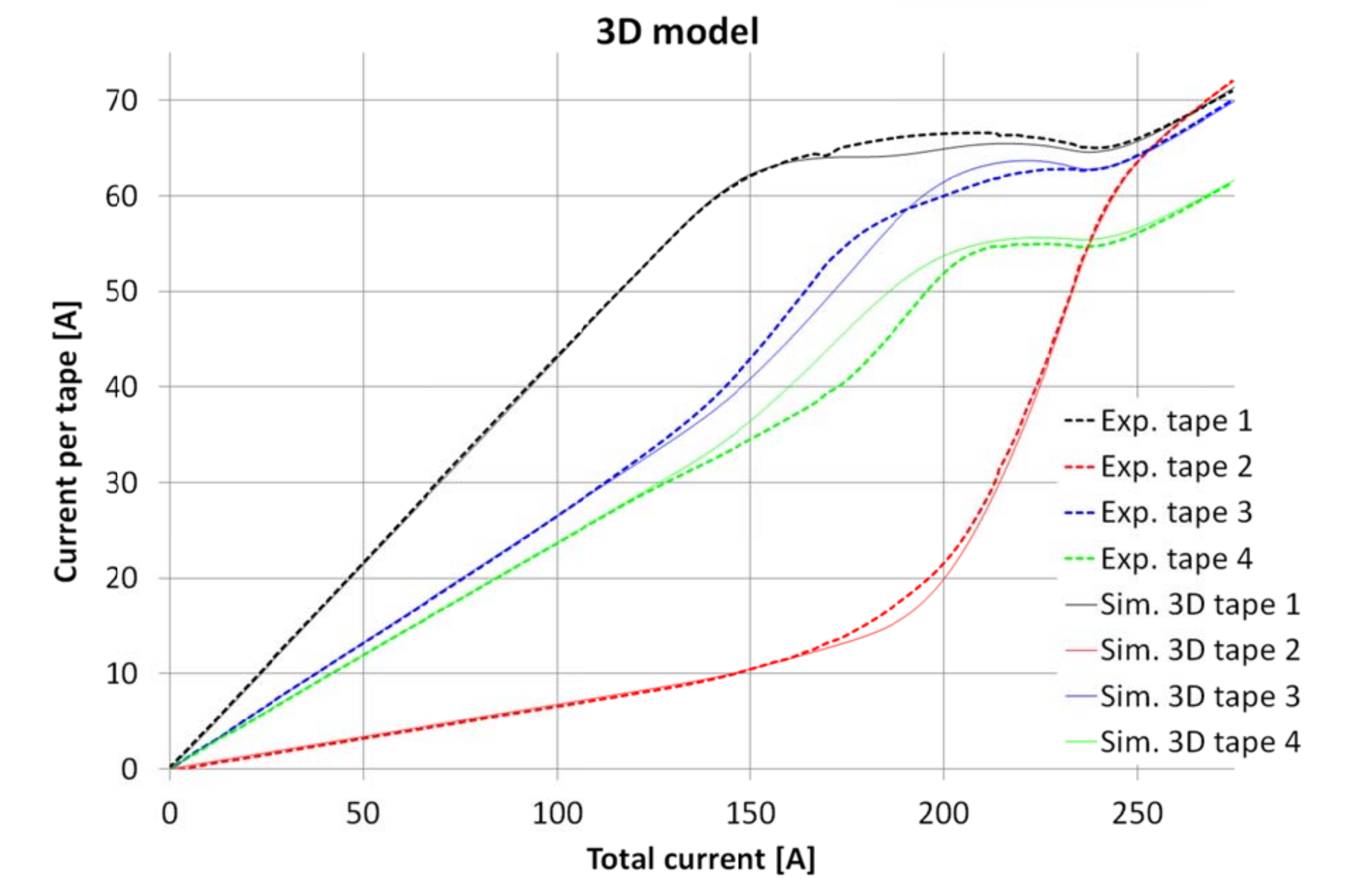

Figure 8 Current per tape vs. total current in the cable as computed by the 3D model. Dashed lines denote the experimental data and solid lines, the simulation results for the respective tapes.

.

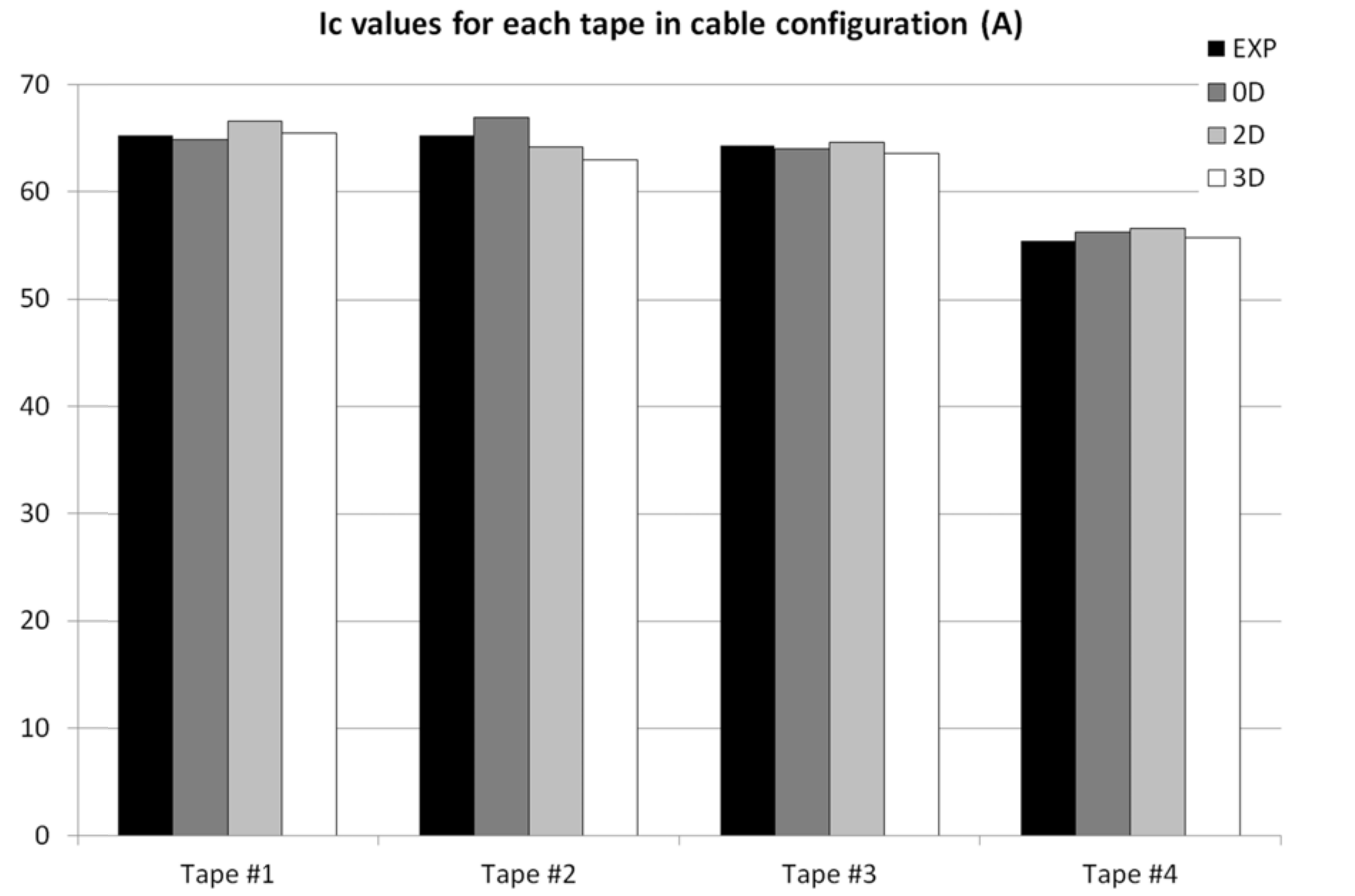

Figure 9 Critical current values for each tape in cable configuration calculated by each model and experimental data.

.

| **Parameter** | **Value** | **Description** |
|---|---|---|
| $I_{c_1}$ | 64.9 A | Critical current of tape 1 |
| $I_{c_2}$ | 67.0 A | Critical current of tape 2 |
| $I_{c_3}$ | 64.1 A | Critical current of tape 3 |
| $I_{c_4}$ | 56.2 A | Critical current of tape 4 |
| $R_1$ | 0.328 μΩ | Termination resistance for tape 1 |
| $R_2$ | 2.101 μΩ | Termination resistance for tape 2 |
| $R_3$ | 0.533 μΩ | Termination resistance for tape 3 |
| $R_4$ | 0.6 μΩ | Termination resistance for tape 4 |
| $n$ | 21 | Exponent of the power law |
| $V_c$ | 60 μV | Critical voltage (60 cm long cable) |

Table 1. Input parameters.